\documentclass[11pt,a4paper]{article}
\pdfoutput=1
\usepackage{jheppub}

\DeclareMathOperator{\re}{Re}
\DeclareMathOperator{\tr}{Tr}
\DeclareMathOperator{\diag}{diag}

\newcommand{\h}{\mathcal H}
\newcommand{\y}{\Upsilon}
\newcommand{\mh}{{\hat m}}
\newcommand{\yh}{{\hat y}}
\newcommand{\xh}{{\hat x}}
\newcommand{\muh}{{\hat \mu}}
\renewcommand{\dh}{{\hat \delta}}
\newcommand{\ddh}{{\hat d}}
\newcommand{\I}{\mathcal{I}}

\title{\boldmath Geometry dependence of RMT-based methods to extract
  the low-energy constants $\Sigma$ and $F$}

\author[a,b]{Christoph Lehner,}
\author[a]{Jacques Bloch,}
\author[c]{Shoji Hashimoto,}
\author[a]{and Tilo Wettig}
\affiliation[a]{Institute for Theoretical Physics, University of
  Regensburg, 93040 Regensburg, Germany}
\affiliation[b]{RIKEN/BNL Research Center, Brookhaven National
  Laboratory, Upton, NY-11973, USA}
\affiliation[c]{High Energy Accelerator Research Organization (KEK),
  Tsukuba 305-0801, Japan}
\emailAdd{clehner@quark.phy.bnl.gov}
\emailAdd{jacques.bloch@physik.uni-regensburg.de}
\emailAdd{shoji.hashimoto@kek.jp}
\emailAdd{tilo.wettig@physik.uni-regensburg.de}

\abstract{The lowest-order low-energy constants $\Sigma$ and $F$ of
  chiral pertubation theory can be extracted from lattice data using
  methods based on the equivalence of random matrix theory (RMT) and
  QCD in the epsilon regime.  We discuss how the choice of the lattice
  geometry affects such methods.  In particular, we show how to
  minimize systematic deviations from RMT by an optimal choice of the
  lattice geometry in the case of two light quark flavors.  We
  illustrate our findings by determining $\Sigma$ and $F$ from lattice
  configurations with two dynamical overlap fermions generated by
  JLQCD, using two different lattice geometries.}

\keywords{epsilon regime, random matrix theory, lattice QCD, imaginary
  chemical potential, geometry dependence, low-energy constants}

\subheader{\footnotesize BNL-94618-2011-JA, KEK-CP-251, RBRC-883}

\begin{document}
\maketitle

\section{Introduction}
It is well known that QCD in a finite volume $V$ at small quark masses
$m$ simplifies as the Compton wavelength of the pion, $m_\pi^{-1}$,
becomes large compared to $V^{1/4}$ \cite{Gasser:1987ah}.  In this
limit the space-time dependence of the low-energy effective theory is
suppressed and the theory is dominated by the constant mode of the
pions.  The distribution of the low-lying eigenvalues of the Dirac
operator can then be calculated in random matrix theory
\cite{Shuryak:1992pi}, see ref.~\cite{Verbaarschot:2000dy} for a
review.  The low-energy constants (LECs) of chiral perturbation theory
are used to map the dimensionful quantities of QCD (or the effective
theory) to the dimensionless quantities of RMT, see, e.g.,
ref.~\cite{Basile:2007ki}.  Matching lattice data for the low-lying
Dirac eigenvalues to RMT results then allows for a determination of
phenomenologically important LECs.

The lowest-order LECs are $\Sigma$ and $F$.  While $\Sigma$ can be
determined rather easily from the distribution of the small Dirac
eigenvalues, $F$ can be determined only if one includes a suitable
constant background gauge field \cite{Sachrajda:2004mi,Mehen:2005fw}
such as isospin imaginary chemical potential
\cite{Damgaard:2005ys,Akemann:2006ru}.  In the following we discuss
the geometry dependence of these methods and show how to minimize
systematic deviations from RMT by an optimal choice of the lattice
geometry.  We also compare our findings with lattice data of the
two-flavor epsilon-regime run of JLQCD
\cite{Fukaya:2007fb,Fukaya:2007yv} and extract $\Sigma$ and $F$ from
these configurations.

The paper is structured as follows.  In section~\ref{sec:nnlo} we
briefly review the epsilon expansion of chiral perturbation theory at
next-to-next-to-leading order (NNLO) which allows for a systematic
discussion of the geometry dependence of RMT-based methods.  In
section~\ref{sec:rmt} we summarize relevant results of RMT for the
distribution of the lowest Dirac eigenvalues at small imaginary
chemical potential.  In section~\ref{sec:num} we compare the analytic
predictions of section~\ref{sec:nnlo} and \ref{sec:rmt} to lattice
data of JLQCD.  We conclude in section~\ref{sec:conc}.

\section{The epsilon expansion at NNLO}\label{sec:nnlo}
In this section we briefly review the epsilon expansion at
NNLO with a small imaginary chemical
potential $i\mu$, see ref.~\cite{Lehner:2010mv}.  In the domain where
the Compton wavelength of the pion becomes large compared to
$V^{1/4}$, chiral perturbation theory ($\chi$PT) can be reordered
according to the power counting \cite{Gasser:1987ah}
\begin{align}
  V \sim \varepsilon^{-4}\,, \qquad
  \partial_\rho \sim \varepsilon\,, \qquad
  \pi(x) \sim \varepsilon\,, \qquad
  {m_\pi} \sim \varepsilon^2\,, \qquad
  \mu \sim \varepsilon^2
\end{align}
with covariant derivative $\partial_\rho$, pion fields $\pi(x)$, pion
mass $m_\pi$, and chemical potential $\mu$.\footnote{In the chiral
  effective theory a nonzero chemical potential is introduced through the
  covariant derivative of the gauged flavor symmetry (see, e.g.,
  ref.~\cite{Lehner:2009pz}), and therefore the power counting of
  $\mu$ is fixed by the power counting of $\partial_\rho$.}  The
corresponding systematic expansion of $\chi$PT is called epsilon
expansion.  To each order in $\varepsilon^2$ one can integrate out the
space-time dependence and obtain a finite-volume effective theory in
terms of the constant pion mode.  The order in $\varepsilon^2$ then
translates into the order in $1/(F^2\sqrt{V})$.  At leading order the
finite-volume effective action is given by
\begin{align}\label{eqn:slo}
  S^\text{LO}_\text{eff} &= -\frac12 V\Sigma\tr( M^\dagger U_0
 + U_0^{-1} M )-\frac12 V F^2 \tr( C
  U_0^{-1} C U_0 )
\end{align}
with constant pion mode
\begin{align}
  U_0 = \exp[i \pi_0] \,, \qquad \pi_0 = \frac1V \int d^4x\: \pi(x)\,,
\end{align}
quark mass matrix $M=\diag(m_1,\ldots,m_{N_f})$, and quark chemical
potential matrix $C=\diag(\mu_1,$ $\ldots,\mu_{N_f})$, where $m_f$ is
the quark mass and $i\mu_f$ is the imaginary chemical potential of
quark flavor $f=1,\ldots,N_f$.  We find that $S_\text{eff}^\text{LO}$
is identical to the RMT action with nonzero chemical potential
\cite{Akemann:2006ru}.  Note that the pion decay constant $F$ drops
out for vanishing chemical potential.  At next-to-leading order (NLO)
in $\varepsilon^2$ the general form of eq.~\eqref{eqn:slo} remains
unchanged with $\Sigma \to \Sigma^\text{NLO}_\text{eff}$, $F \to
F^\text{NLO}_\text{eff}$, see
refs.~\cite{Damgaard:2007xg,Akemann:2008vp,Lehner:2009pz} for explicit
expressions.  In an actual lattice simulation we measure effective
values $\Sigma_\text{eff}$ and $F_\text{eff}$, and we need to include
finite-volume corrections to recover the infinite-volume values
$\Sigma$ and $F$.

At NNLO and to leading order in the small chemical
potential\footnote{There are also NNLO terms proportional to $V^2C^4$
  that have been omitted in \eqref{eqn:snnlo}.} the effective action
has the form \cite{Lehner:2010mv}
\begin{align}\label{eqn:snnlo}
  S^\text{NNLO}_\text{eff} &= -\frac12 V\Sigma^\text{NNLO}_\text{eff}\tr( M^\dagger U_0
  \notag + U_0^{-1} M )-\frac12  V (F^\text{NNLO}_\text{eff})^2 \tr( C
  U_0^{-1} C U_0 )\notag\\&\quad
 +\y_1
  \Sigma (VF)^2   \tr(C)[\tr(U_0\{M^\dagger, C\})+\tr(U_0^{-1}\{C,M\})]
  \notag\\&\quad
  +\y_2 \Sigma (VF)^2\tr(\{M^\dagger, C\}U_0 C + \{C, M\} C U_0^{-1}
  \notag\\&\qquad\qquad\qquad\qquad + \{U_0, C\} U_0^{-1} C U_0
  M^\dagger + CU_0 \{C,U_0^{-1}\} M U_0^{-1} ) \notag\\&\quad
 +\y_3
  \Sigma (VF)^2 \tr(U_0^{-1}CU_0C + C^2) \tr(M U_0^{-1} +
  M^\dagger U_0) \notag\\&\quad
 +\y_4 \Sigma (VF)^2 \tr(U_0^{-1}CU_0C - C^2) \tr(M
  U_0^{-1} + M^\dagger U_0)\notag \\&\quad +\y_5 \Sigma (VF)^2\tr(
  [M^\dagger, C] U_0 C + [C,M]C U_0^{-1} \notag\\&\qquad\qquad\qquad\qquad
 + [U_0, C] U_0^{-1}C U_0 M^\dagger
  + CU_0 [C,U_0^{-1}] M U_0^{-1} ) \notag\\&\quad+\y_6
  (V\Sigma)^2 [\tr(M U_0^{-1}+M^\dagger U_0)]^2  +\y_7
  (V\Sigma)^2 [\tr(M U_0^{-1} - M^\dagger U_0)]^2 \notag\\&\quad +\y_8
  (V\Sigma)^2 [\tr(M U_0^{-1}MU_0^{-1})+\tr(M^\dagger U_0M^\dagger
  U_0)] \notag\\&\quad +\h_1VF^2\tr( C^2 )
  +\h_2 (V\Sigma)^2 \tr(M^\dagger M) + \h_3 V F^2 (\tr C)^2
\end{align}
with finite-volume effective coupling constants $\y_i$ and
$\h_i$.
The LECs $\Sigma$ and $F$ also receive further corrections, $\Sigma
\to \Sigma^\text{NNLO}_\text{eff}$ and $F \to
F^\text{NNLO}_\text{eff}$.  The terms in eq.~\eqref{eqn:snnlo} that
were not present in eq.~\eqref{eqn:slo} cannot be mapped to RMT.
These terms are proportional to the $\y_i$ and $\h_i$.  Therefore the
magnitude of these coefficients determines the systematic deviations
from RMT of, e.g., Dirac eigenvalue distributions.  The coefficients
$\h_1$ and $\h_3$ do not couple to $U_0$ or $M$ and are therefore
irrelevant for Dirac eigenvalue distributions (which involve
derivatives with respect to $M$ in the partially quenched theory).  
The coefficients
$\y_i$, $\h_2$, $\Sigma^\text{NNLO}_\text{eff}$, and
$F^\text{NNLO}_\text{eff}$ depend on the NLO LECs of $\chi$PT and
on the geometry of the space-time box through finite-volume
propagators.  Explicit results are given in \cite{Lehner:2010mv}.

\begin{figure}[t]
  \centering
  \includegraphics{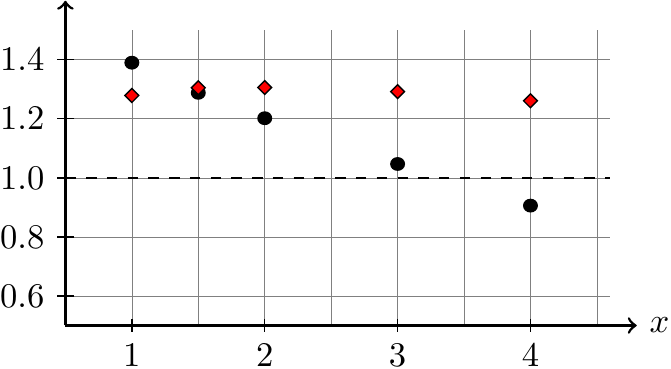} \hspace{.5cm}
  \includegraphics{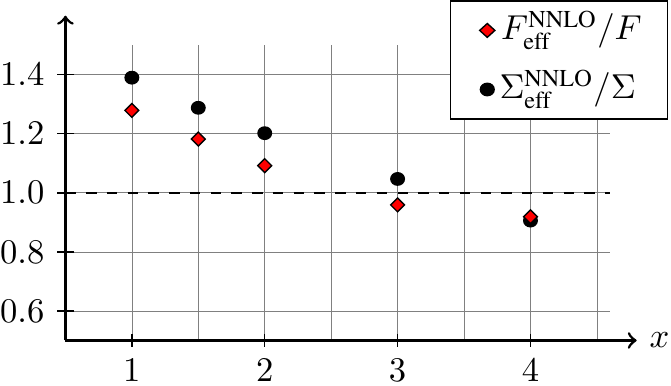}
  \caption{Finite-volume corrections to $\Sigma$ and $F$ for
    geometries $(a_x)$ on the left and $(b_x)$ on the right with
    parameters $F=90$ MeV, $L=1.71$ fm, and $m_\pi^2\sqrt{V}=1$.
    Taken from ref.~\cite{Lehner:2010mv}.}
  \label{fig:effnnlo}
\end{figure}

To be specific we discuss the following lattice geometries from now on,
\begin{subequations}
  \begin{align}
    (a_x)\qquad {L}_0 &= x L\,,\qquad {L}_1 = {L}_2 = {L}_3 = L\,, \\
    (b_x)\qquad {L}_3 &= x L\,,\qquad {L}_0 = {L}_1 = {L}_2 = L\,,
  \end{align}
\end{subequations}
where $x \in \{1, 3/2, 2, 3, 4\}$, and $L_i$ is the extent of the
space-time box in direction $i$ ($i=0$ denotes the temporal direction
to which $\mu$ couples).  In figure~\ref{fig:effnnlo} we show the
finite-volume corrections to $\Sigma$ and $F$ for the different
geometries at NNLO for a set of parameters similar to the parameters
of the JLQCD two-flavor epsilon-regime run
\cite{Fukaya:2007fb,Fukaya:2007yv}.
We note that the finite-volume corrections to $\Sigma$ are invariant
under $(a_x)\leftrightarrow(b_x)$, while the finite-volume corrections
to $F$ depend on the choice of geometry.  The reason is that the
permutation symmetry of the four space-time dimensions is broken by
the chemical potential, to which $F$ couples.  For our choice of
parameters, geometry $(b_x)$ leads to smaller finite-volume
corrections to $F$ than geometry $(a_x)$.  This was also observed in
ref.~\cite{Lehner:2009pz} at NLO.

We continue our discussion with the finite-volume effective coupling
constants $\y_i$ and $\h_2$ that are responsible for the systematic
deviations from RMT.  The high-energy constant $\h_2$ should not
contribute to low-energy phenomenology (as was shown explicitly for
the spectral density in \cite{Damgaard:2008zs}), and therefore we
do not discuss it further.  It is an interesting observation
\cite{Lehner:2010mv} that $\y_1,\y_2,\y_3$ do not depend on the NLO
LECs of $\chi$PT and depend on the geometry only through a common
coefficient $\gamma$, i.e.,
\begin{align}\label{eqn:gamma}
    \y_1,\y_2,\y_3 \propto \gamma\,.
\end{align}
The coefficient $\gamma$ changes under $(a_x)\leftrightarrow(b_x)$,
while $\y_4,\ldots,\y_8$ (and $\h_2$) are invariant under the same
exchange \cite{Lehner:2010mv}.  This implies that for nonzero chemical
potential a judicious choice of geometry is possible which minimizes
the systematic deviations from RMT due to $\y_1,\y_2,\y_3$ at a given
volume.  Of course, for zero chemical potential $(a_x)$ and $(b_x)$
are equivalent, so the choice of geometry has no impact.

\begin{figure}[t]
  \centering
  \includegraphics{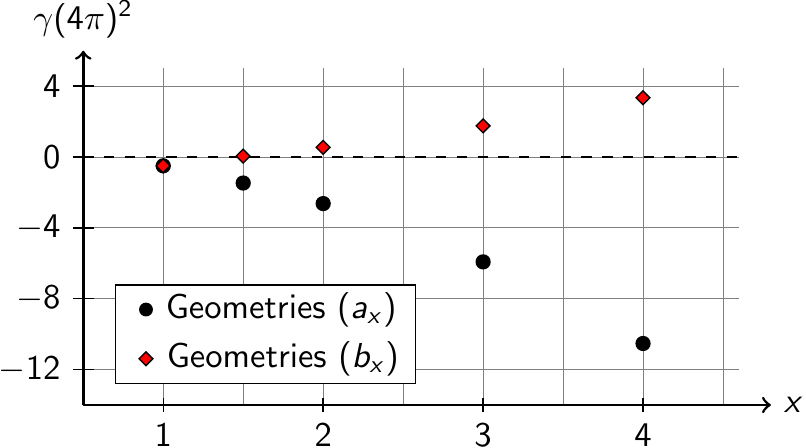}
  \caption{Geometry dependence of systematic deviations from RMT.
    Taken from ref.~\cite{Lehner:2010mv}.}
  \label{fig:nonuniv}
\end{figure}

We plot $\gamma$ for different geometries in
figure~\ref{fig:nonuniv} for the same set of parameters used in
figure~\ref{fig:effnnlo}.\footnote{Since the $\Upsilon_i$ are of order
$1/(4\pi)^2$ we plot $\gamma(4\pi)^2$, which is then of order 1.}
We note that the coefficient $\gamma$, and thus a part of the
systematic deviations from RMT, can be reduced significantly by
choosing the geometry $(b_x)$ instead of $(a_x)$ for the same value of
the asymmetry $x$.  Explicit numbers for the $\Upsilon_i$ and their
impact on systematic errors will be discussed in Sec.~\ref{sec:num}.
Note that for large asymmetries $x$ the coefficients $\Upsilon_i$ grow
rapidly.  Thus for too large values of $x$ the epsilon expansion
breaks down.  This corresponds to the largest individual dimension
$\max(L_i)$ being significantly larger than the Compton wavelength of
the pion.

Note that we can project out a single topological sector by modifying
the integration domain of the constant pion mode $U_0$ and including a
volume-independent determinant term in the partition function, see,
e.g., refs.~\cite{Verbaarschot:2000dy,Lehner:2009pz}.  Therefore the
discussion of systematic deviations from RMT presented in this section
is also valid for fixed topological charge.

\section{Random matrix theory}\label{sec:rmt}
In this section we summarize some important results of random matrix
theory that can be used to determine $\Sigma$ and $F$ from fits to
Dirac eigenvalue distributions.  We consider chiral random matrix
theory with imaginary chemical potential defined by the partition
function
\begin{align}
  \label{eqn:rmtdef1}
  Z_\nu = \int dV dW \, e^{-N \tr(W^\dagger W + V^\dagger V)}
  \prod_{f=1}^{N_f} \det(D(\mu^r_f)+m^r_f)\,,
\end{align}
where $m^r_1$, $\ldots$, $m^r_{N_f}$ ($i\mu^r_1$, $\ldots$,
$i\mu^r_{N_f}$) are the masses (imaginary chemical potentials) of the
sea quarks, see refs.~\cite{Akemann:2006ru,Osborn:2004rf}, the latter
for the case of real chemical potential.  The integral is over the
real and imaginary parts of the elements of the complex $N \times (N +
\nu)$ matrices $W$ and $V$ with Cartesian integration measure.  The
random matrix Dirac operator is defined by
\begin{align}
  \label{eqn:rmtdef2}
  D(\mu^r_f) =
  \begin{pmatrix}
    0 & i V + i\mu^r_f W \\
    i V^\dagger + i\mu^r_f W^\dagger & 0
  \end{pmatrix},
\end{align}
which has $|\nu|$ eigenvalues equal to zero.  Therefore $\nu$ is
interpreted as the topological charge.  Note that $m^r_f$ and
$\mu^r_f$ are dimensionless quantities.  They have to be mapped to
physical quantities by comparison with the low-energy effective theory
of QCD.  It was shown in ref.~\cite{Basile:2007ki} that in the limit
$N \to \infty$ chiral random matrix theory can be mapped to chiral
perturbation theory using
\begin{align}\label{eqn:rmtmatch}
  \mh_f = m_f V \Sigma = 2 N m^r_f\,,\qquad
  \muh_f^2 = \mu_f^2 F^2 V = 2 N (\mu_f^r)^2\,,
\end{align}
where $f$ denotes an arbitrary quark flavor, $m_f$ is the physical
quark mass, and $\mu_f$ is the physical chemical potential.  Thus, the
low-energy constants $\Sigma$ and $F$ appear in the conversion from
physical units to dimensionless random matrix units.  Note that
refs.~\cite{Basile:2007ki} and \cite{Akemann:2006ru} use a different
notation for the dimension of the random matrix Dirac operator.  The
quantities $\mh_f$ and $\muh_f$ are often referred to as microscopic
scaling quantities due to the limit $N \to \infty$.

The eigenvalue correlation functions for the random matrix model
defined by eqs.~\eqref{eqn:rmtdef1} and \eqref{eqn:rmtdef2} in the
limit of $N\to\infty$ were calculated in ref.~\cite{Akemann:2006ru}.
In this section we consider the case of $N_f=2$ sea quarks with masses
$\mh_u$ and $\mh_d$ at zero chemical potential.  This setup
corresponds to the two-flavor simulation of JLQCD
\cite{Fukaya:2007fb,Fukaya:2007yv} that is described in more detail in
section~\ref{sec:num}.  We then compute Dirac eigenvalues $\xh$ at
zero chemical potential and $\yh$ at imaginary chemical potential
$i\dh$.  Note that we could equally well have used a setup with $\xh$
at imaginary chemical potential $-i\dh/2$ and $\yh$ at imaginary
chemical potential $+i\dh/2$ since only the isospin component of the
chemical potential is relevant for eigenvalue correlation functions
\cite{Akemann:2006ru}.

We define the two-point correlator
\begin{align}\label{eqn:twocorrrmt}
  \rho^{(2)}_{(1,1)}(\xh,\yh) =
  \bigg\langle\sum_{n,m}\delta(\xh-\hat\lambda_n(\muh=0))
  \delta(\yh-\hat\lambda_m(\muh=\delta))\bigg\rangle\,,
\end{align}
where $\hat\lambda_n= 2N\lambda^r_n=\lambda_n V \Sigma$ and the sum is
over all eigenvalues $\lambda^r_n$ of the random matrix Dirac operator
at chemical potential $\muh=0$ and $i\muh=i\dh$.  This correlator allows
for a discussion of the shift of Dirac eigenvalues due to the
imaginary chemical potential $i\dh$.  Equation~\eqref{eqn:twocorrrmt}
is calculated in ref.~\cite{Akemann:2006ru}.  The result is given by
\begin{align}
  \rho_{(1,1)}^{(2)} (\xh,\yh) = \xh\, \yh\: & \det
  \begin{bmatrix}
    J_\nu(i \mh_u) & i \mh_u J_{\nu+1}(i\mh_u) \\
    J_\nu(i \mh_d) & i \mh_d J_{\nu+1}(i\mh_d)
  \end{bmatrix}^{-2} \det \begin{bmatrix}
    \Psi_{11} & \Psi_{12} \\
    \Psi_{21} & \Psi_{22}
  \end{bmatrix},
\end{align}
where $J_\nu$ is the Bessel function of the first kind,
\begin{subequations}
  \begin{align}
    \Psi_{11} &= \det\begin{bmatrix}
      \I^0(\xh,i\mh_u) & J_\nu(i\mh_u) & i\mh_u J_{\nu+1}(i\mh_u) \\
      \I^0(\xh,i\mh_d) & J_\nu(i\mh_d) & i\mh_d J_{\nu+1}(i\mh_d) \\
      \I^0(\xh,\xh) & J_\nu(\xh) & \xh J_{\nu+1}(\xh)
    \end{bmatrix}, \\
    \Psi_{12} &= \det\begin{bmatrix}
      \I^0(\xh,i\mh_u) & J_\nu(i\mh_u) & i\mh_u J_{\nu+1}(i\mh_u) \\
      \I^0(\xh,i\mh_d) & J_\nu(i\mh_d) & i\mh_d J_{\nu+1}(i\mh_d) \\
      -\tilde \I^-(\xh,\yh) & e^{-\dh^2/2} J_\nu(\yh) & e^{-\dh^2/2} G_\nu(\yh,\dh)
    \end{bmatrix}, \\
    \Psi_{21} &= \det\begin{bmatrix}
      \I^+(\yh,i\mh_u) & J_\nu(i\mh_u) & i\mh_u J_{\nu+1}(i\mh_u) \\
      \I^+(\yh,i\mh_d) & J_\nu(i\mh_d) & i\mh_d J_{\nu+1}(i\mh_d) \\
      \I^+(\yh,\xh) & J_\nu(\xh) & \xh J_{\nu+1}(\xh)
    \end{bmatrix}, \\
    \Psi_{22} &= \det\begin{bmatrix}
      \I^+(\yh,i\mh_u) & J_\nu(i\mh_u) & i\mh_u J_{\nu+1}(i\mh_u) \\
      \I^+(\yh,i\mh_d) & J_\nu(i\mh_d) & i\mh_d J_{\nu+1}(i\mh_d) \\
      \I^0(\yh,\yh) & e^{-\dh^2/2} J_\nu(\yh) & e^{-\dh^2/2} G_\nu(\yh,\dh)
    \end{bmatrix}
  \end{align}
\end{subequations}
with
\begin{subequations}
  \begin{align}
    \I^0(\xh,\yh) & = \frac 12 \int_0^1 dt\: J_\nu(\xh \sqrt t) J_\nu(\yh \sqrt t) 
    = \frac{\xh J_{\nu+1}(\xh) J_\nu(\yh) - \yh J_{\nu+1}(\yh) J_\nu(\xh)}{\xh^2 - \yh^2}\,,\\
    \I^\pm(\xh,\yh) & = \frac 12 \int_0^1 dt\: e^{\pm \dh^2 t/2} J_\nu(\xh \sqrt t) J_\nu(\yh \sqrt t)\,, \\
    \tilde \I^-(\xh,\yh) & = \frac 1 {\dh^2} \exp\left( - \frac{\xh^2 + \yh^2}{2 \dh^2} \right) 
    I_\nu\left(\frac{\xh \yh}{\dh^2}\right) - \I^-(\xh,\yh)\,, \\
    G_\nu(\yh,\dh) & = \yh J_{\nu+1}(\yh) + \dh^2 J_\nu(\yh)\,,
  \end{align}
\end{subequations}
and  $I_\nu$ is the modified Bessel function.

In the limit of small chemical potential $\dh^2 \ll 1$ the term
proportional to $\dh^{-2}$ in $\tilde \I^-$ dominates.  Furthermore,
we can perform a large-argument expansion of the Bessel function in
$\tilde \I^-$ and ignore all terms of order $\dh^2$, so that
\begin{align}
  \rho_{(1,1)}^{(2)} (\xh,\yh) & = H_\nu(\xh,\yh,\mh_u,\mh_d)
  \frac1{\sqrt{2 \pi \dh^2}} \exp\left( - \frac{(\xh - \yh)^2}{2
      \dh^2} \right)
\end{align}
with
\begin{align}
  H_\nu(\xh,\yh,\mh_u,\mh_d) &=  \sqrt{\xh \yh} \:\frac{\det\begin{bmatrix}
      \I^0(\yh,i\mh_u) & J_\nu(i\mh_u) & i\mh_u J_{\nu+1}(i\mh_u) \\
      \I^0(\yh,i\mh_d) & J_\nu(i\mh_d) & i\mh_d J_{\nu+1}(i\mh_d) \\
      \I^0(\yh,\xh) & J_\nu(\xh) & \xh J_{\nu+1}(\xh)
    \end{bmatrix}}{\det
    \begin{bmatrix}
      J_\nu(i \mh_u) & i \mh_u J_{\nu+1}(i\mh_u) \\
      J_\nu(i \mh_d) & i \mh_d J_{\nu+1}(i\mh_d)
    \end{bmatrix}}\,.
\end{align}
Note that the prefactor $H_\nu$ is independent of $\dh$.  Let us
define a probability distribution that measures the shift $\ddh$ of
the eigenvalues due to the imaginary chemical potential $i\dh$ up to a
cutoff $\xh_c$,
\begin{align}
  P(\ddh,\xh_c) & = \frac1{{\cal N}(x_c)}\int_0^{\xh_c} d\xh\:
  \rho_{(1,1)}^{(2)}(\xh,\xh+\ddh)
  \notag\\
  & = \tilde H_\nu(\ddh,\xh_c,\mh_u,\mh_d)\frac1{\sqrt{2 \pi \dh^2}}
  \exp\left( - \frac{\ddh^2}{2 \dh^2} \right)
\end{align}
with
\begin{align}
  \tilde H_\nu(\ddh,\xh_c,\mh_u,\mh_d) &= \frac1{{\cal
      N}(x_c)}\int_0^{\xh_c} d\xh\: H_\nu(\xh,\xh+\ddh,\mh_u,\mh_d)\,, \notag\\
  {\cal N}(x_c) &= \int d\ddh \int_0^{\xh_c} d\xh\:
  \rho_{(1,1)}^{(2)}(\xh,\xh+\ddh)\,.
\end{align}
The Gaussian factor peaks strongly at $\ddh=0$, and thus we can expand
$\tilde H_\nu$ about $\ddh=0$ to linear order in $\ddh$.  The constant
term in the expansion is fixed by the normalization
\begin{align}
  \int d\ddh\: P(\ddh,\xh_c) = 1  
\end{align}
for $\dh^2 \to 0$.  Therefore we have
\begin{align}
  P(\ddh,\xh_c) & = \frac1 {\sqrt{2 \pi \dh^2}} \exp\left(
    -\frac{\ddh^2}{2 \dh^2} \right) (1 + c_1 \ddh + {\cal O}(\ddh^2)) \,,
\end{align}
where only $c_1$ depends on $\xh_c$.  Note that to first order in
$\ddh$, $P(\ddh,\xh_c)$ corresponds to a Gaussian distribution with
width $\dh$ and center $c_1 \dh^2$.

In section~\ref{sec:num} we use a small $\dh$ for the numerical fits.
Due to the Gaussian factor, $\ddh$ is of order $\dh$, and therefore
the contribution of $c_1$ can be neglected (we have confirmed this for
our numerical results in section~\ref{sec:num}).  We define
$P(\ddh)$ to be the leading contribution to
$P(\ddh,\xh_c)$ in the limit of small $\dh$, and therefore
\begin{align}\label{eqn:defpd}
  P(\ddh) & = \frac 1{\sqrt{2 \pi \dh^2}} \exp\left( -
    \frac{\ddh^2}{2\dh^2}\right).
\end{align}
This quantity is well-suited to determine $\dh$ and therefore $F$ from
a fit to eigenvalue spectra obtained in lattice QCD simulations.  Note
that in the limit of small $\dh$ the distribution does not depend on
$c_1$.  For a related discussion with imaginary isospin chemical
potential we refer to ref.~\cite{Damgaard:2006rh}.

\begin{figure}[t]
  \centering
  \includegraphics{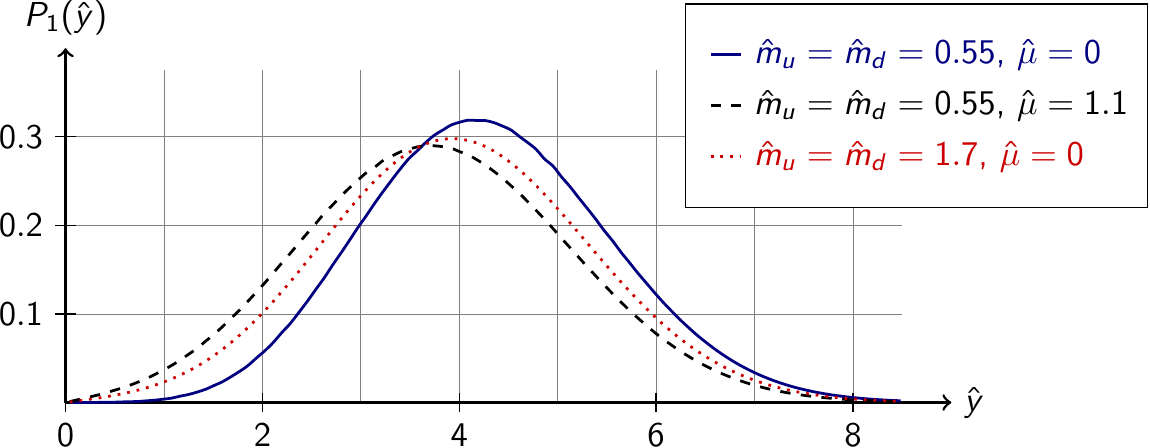}
  \caption{Distribution of the lowest Dirac eigenvalue for different
    quark masses $\mh_u=\mh_d$ and imaginary chemical potentials $i\muh$.}
  \label{fig:lowev}
\end{figure}

In refs.~\cite{Wilke:1997gf,Nishigaki:1998is} the distribution of the
lowest Dirac eigenvalue $\yh$ was calculated analytically, and in
ref.~\cite{Akemann:2008va} the calculation was extended to nonzero
imaginary chemical potential $i\muh$.  We use the notation of
ref.~\cite{Akemann:2008va}.  The distribution of the lowest eigenvalue
is given by
\begin{align}
  P_1(\yh) = - \partial_{\yh} E_{0,0}^{(0+2)} (\yh,0)
\end{align}
with gap probability
\begin{align}
  E_{0,0}^{(0+2)} (\yh,0) &=
  \frac{2\det
    \begin{bmatrix}
      &Q_S(\yh,\mh_u;t=1) & \partial_t Q_S(\yh,\mh_u;t) \vert_{t=1}& \\
      &Q_S(\yh,\mh_d;t=1) & \partial_t Q_S(\yh,\mh_d;t) \vert_{t=1}&
    \end{bmatrix}
  }{\mh_d I_0(\mh_u) I_1(\mh_d) - \mh_u I_0(\mh_d) I_1(\mh_u)}  \exp\left(-\frac14 \yh^2 - \dh^2 \right),
\end{align}
where
\begin{align}
  Q_S(\yh,\mh;t) &= \frac 12 \int_0^1 dr\: e^{r (t/2) \dh^2}
  I_0(\sqrt{rt}\: \mh) \sqrt{\frac{t}{1-r}}\: \yh\: I_1(\sqrt{(1-r)t}\: \yh)
  + e^{(t/2) \dh^2} I_0(\sqrt{t}\: \mh)\,.
\end{align}
In figure~\ref{fig:lowev} we display $P_1(\yh)$ for different values
of $\muh$ and $\mh_u=\mh_d$.  Note that the dependence on $\muh$ and
$\mh_u$ is strongly correlated, and therefore it is challenging to use
this quantity to determine both $\Sigma$ and $F$ from a fit to
numerical data.  Nevertheless, the distribution of the lowest
eigenvalue is well-suited to determine the scale of $\yh$ for $\muh=0$
and therefore $\Sigma$.

\section{Numerical results}\label{sec:num}
In this section we check the results of section \ref{sec:nnlo} against
the epsilon-regime run of JLQCD with two dynamical overlap fermions
with masses $am_u = am_d = 0.002$ and $16^3\times32$ lattice points at
lattice spacing $a = 0.1091(23)$ fm
\cite{Fukaya:2007fb,Fukaya:2007yv}.  For these parameters we have
$m_\pi\min(L_i) \simeq 1$, $m_\pi\max(L_i) \simeq 2$, and $m_\pi^2
\sqrt V \simeq 1.34$ (using the GOR value $m_\pi=110\text{ MeV}$, see
below).  The sea quarks are at zero chemical potential, and topology
is fixed to $\nu=0$.  We compute the eigenvalues of the valence
overlap Dirac operator on 460 configurations at zero and nonzero
imaginary chemical potential.\footnote{In order to introduce a nonzero
  imaginary chemical potential $i\mu$ in the overlap Dirac operator we
  multiply the forward (backward) temporal links by a factor of
  $e^{ia\mu}$ ($e^{-ia\mu}$), see \cite{Bloch:2006cd}.}  In this way
the existing configurations can be used to extract $\Sigma$ and $F$
with low numerical cost.

\begin{figure}[t]
  \centering
  \includegraphics{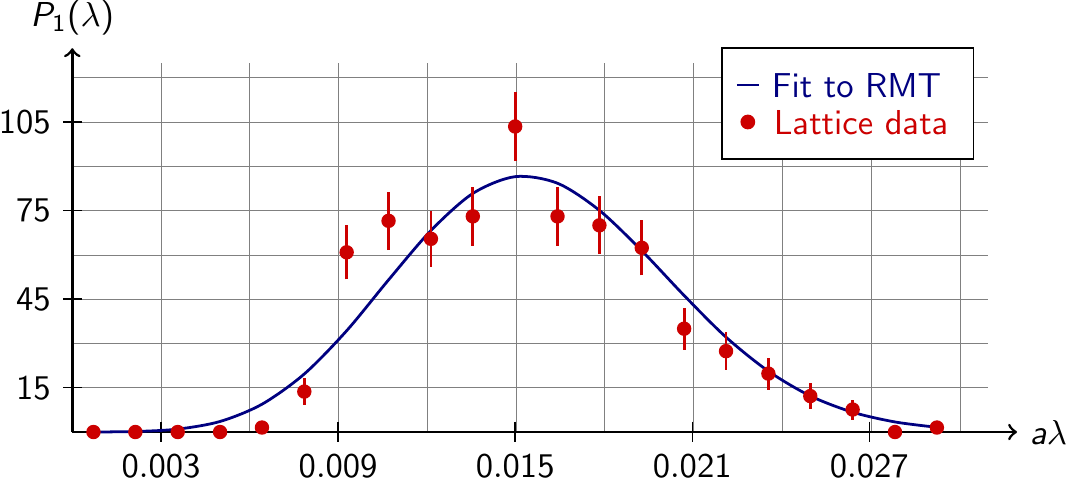}    
  \caption{Fit to lowest-lying Dirac eigenvalue distribution
    $P_1(\lambda)$ with $\chi^2 / \text{dof}={2.9}$,
    $a^3\Sigma_\text{eff}=0.00208({2})$.}
  \label{fig:numfevB}
\end{figure}

We first fit the distribution $P_1(\lambda)$ of the lowest-lying Dirac
eigenvalue at zero chemical potential in figure~\ref{fig:numfevB}
in order to extract the finite-volume effective value
$a^3\Sigma_\text{eff} = 0.00208(2)$, where we cite the statistical error.
This corresponds to the dimensionful value
\begin{align}
 \Sigma_\text{eff} = (231(1)(5) \text{ MeV})^3\,,
\end{align}
where we cite the statistical error (left) and the error propagated
from the uncertainty in the lattice spacing (right).  The
dimensionless value is compatible with $a^3\Sigma_\text{eff} =
0.00212(6)$ obtained in ref.~\cite{Fukaya:2007fb} on the same
configurations by a fit to the integrated Dirac eigenvalue
distribution.  Note that $\Sigma$ is renormalization scheme dependent
and that we give only the values for the lattice renormalization
scheme here.  Including finite-volume corrections at NLO gives an
infinite-volume value
\begin{equation}
  \label{eq:Sigma}
  \Sigma=\Sigma_\text{eff}/1.1454=(221(1)(5)\text{ MeV})^3\,.  
\end{equation}
We use only NLO finite-volume corrections here and in the remainder of
this paper since the NNLO finite-volume corrections and the systematic
deviations from RMT are of the same order in $1/F^2\sqrt{V}$.  From
figure~\ref{fig:numfevB} we see that the systematic deviations from
RMT are significant.  As discussed in section~\ref{sec:nnlo}, they
cannot be minimized by a judicious choice of lattice geometry.  To
further reduce the systematic errors in the fit for $\Sigma$, one
would have to go to a larger volume, which is beyond the scope of this
work.  Alternatively, one could compute $P_1(\lambda)$ to NNLO,
including all non-universal terms.  This is a a very difficult
calculation that nobody has attempted yet.

\begin{figure}[t]
  \centering
  \includegraphics{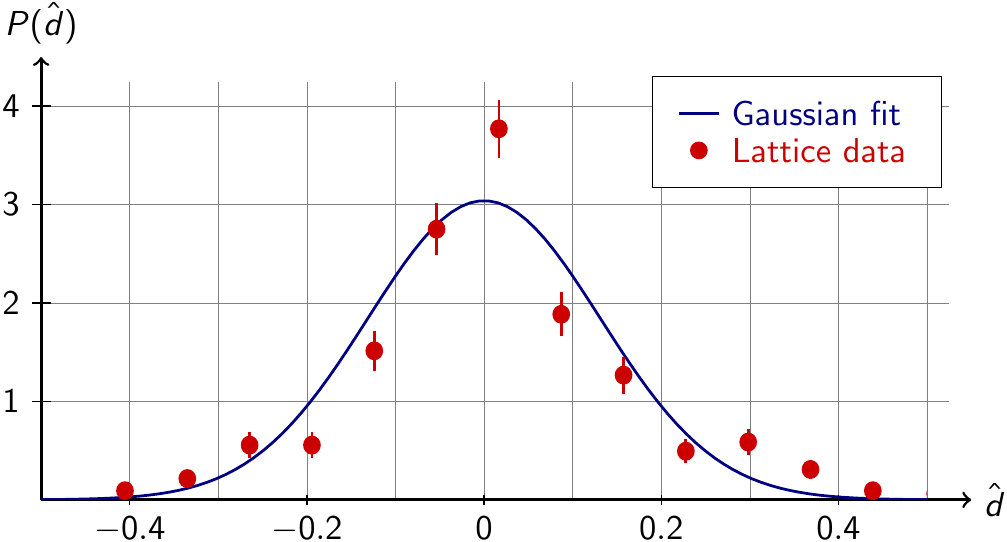}
  \caption{Fit to the distribution of Dirac eigenvalue shifts $P(\hat
    d)$ due to imaginary chemical potential $a\mu=0.01$ with $\hat
    d=d\Sigma V$ in geometry $(a_2)$.  The result is given by
    $F_\text{eff} = 66(5)(1) \text{ MeV}$ with $\chi^2 /
    \text{dof}=4.2$.  We cite the statistical error (left) as well as
    the error propagated from the uncertainty in the lattice spacing
    (right).}
  \label{fig:pda}
\end{figure}

Next we fit the shift of the lowest-lying Dirac eigenvalue due to a
small imaginary chemical potential $i\mu$ in order to extract $F$ as
proposed in ref.~\cite{Damgaard:2005ys}.  As shown in section
\ref{sec:rmt}, RMT predicts a Gaussian distribution with
$\sigma^2=\mu^2 F^2 V$ for the distribution $P$ of the difference
$d$ between the lowest Dirac eigenvalue at zero and at nonzero
imaginary chemical potential, see also
refs.~\cite{Damgaard:2005ys,Akemann:2006ru,Damgaard:2006rh}.  In
figure~\ref{fig:pda} we show the resulting fit for geometry $(a_2)$
with finite-volume effective value
\begin{align}
  F_\text{eff}^{(a_2)} = 66(5)(1) \text{ MeV}\,,
\end{align}
where we cite the statistical error (left) as well as the error
propagated from the uncertainty in the lattice spacing (right).  We
note that the quality of the fit is rather bad
($\chi^2/\text{dof}=4.2$) and that this value is not compatible with
the result from a fit to meson correlators obtained on the same
configurations \cite{Fukaya:2007pn}, $F_\text{meson} = 87.3(5.6)
\text{ MeV}$.  If we include finite-volume corrections at NLO we
obtain the infinite-volume value
\begin{align}
  F^{(a_2)} = {50(4)(1)} \text{ MeV}
\end{align}
so that the agreement is even worse.  The bad $\chi^2/\text{dof}=4.2$
suggests that the non-universal terms at NNLO, see
eq.~\eqref{eqn:snnlo} and the subsequent discussion, affect the
distribution in a non-trivial manner.

\begin{figure}[t]
  \centering
  \includegraphics{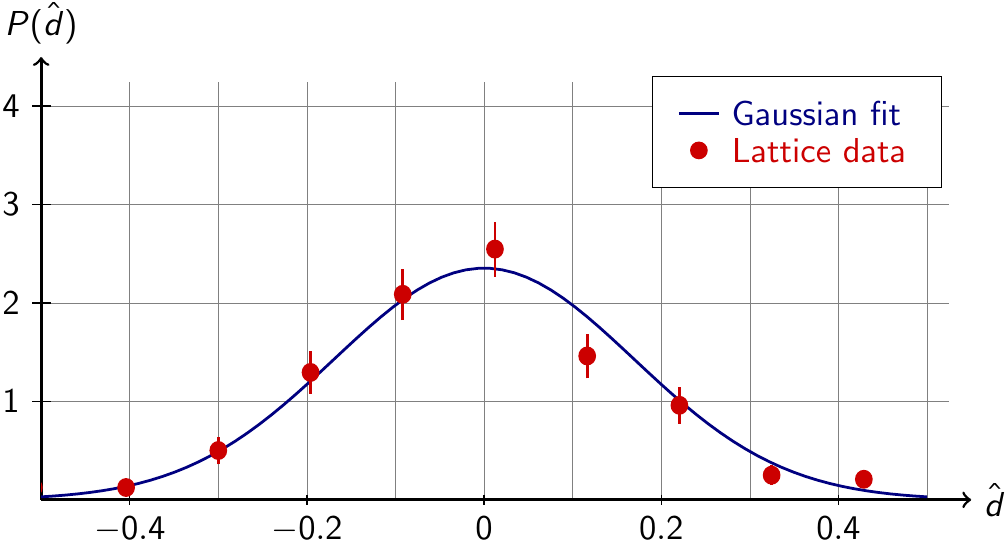}
  \caption{Fit to the distribution of Dirac eigenvalue shifts $P(\hat
    d)$ due to imaginary chemical potential $a\mu=0.01$ with $\hat
    d=d\Sigma V$ in geometry $(b_2)$.  The result is given by
    $F_\text{eff} = 85(5)(2) \text{ MeV}$ with $\chi^2 /
    \text{dof}=0.91$.  We cite the statistical error (left) as well as
    the error propagated from the uncertainty in the lattice spacing
    (right).}
  \label{fig:pdb}
\end{figure}

From our discussion in section~\ref{sec:nnlo} we learned that we can
significantly reduce these systematic deviations from RMT by choosing
lattice geometry $(b_2)$ instead of $(a_2)$.  In practice this means
that we should rotate the lattice by 90 degrees so that we have one
large spatial dimension instead of a large temporal dimension.  In
figure~\ref{fig:pdb} we show the resulting fit for geometry $(b_2)$
with good $\chi^2 / \text{dof}=0.91$ and
\begin{align}\label{eqn:Fbtwo}
  F_\text{eff}^{(b_2)} = 85(5)(2) \text{ MeV}\,.
\end{align}
Including finite-volume corrections at NLO\footnote{If we use
  finite-volume corrections at NNLO the value is further reduced by
  2\%.} this gives
\begin{align}
  \label{eq:Ffinal}
  F^{(b_2)} = 80(5)(2) \text{ MeV}\,,
\end{align}
which agrees within errors with the result from the fit to meson
correlators given above.

The value of $\chi^2 / \text{dof}$ in our fits is stable under
variations of the bin size.  Specifically, if we keep the value of $F$
fixed at the result of eq.~\eqref{eqn:Fbtwo} and vary the bin size
from $0.05$ to $0.25$, the value of $\chi^2 / \text{dof}$ only changes
within about one standard deviation of the $\chi^2 / \text{dof}$
distribution, as theoretically expected.  Based on the good $\chi^2 /
\text{dof}$ in geometry $(b_2)$, we make the hypothesis that in this
geometry the non-universal terms in eq.~\eqref{eqn:snnlo} are small
compared to the universal terms.  Using the Cram\'er--von-Mises
criterion to estimate the goodness of our fit in geometry $(b_2)$ we
obtain a $T$-value of $0.097$, which implies that there is no reason
to reject our hypothesis.

Nevertheless, it is instructive to compute the actual values of the
$\Upsilon_i$ which determine the size of the systematic errors.  One
can show that for $N_f=2$ the traces of the flavor matrices multiplied
by $\Upsilon_4$ and $\Upsilon_5$ in \eqref{eqn:snnlo} are identical,
and therefore \eqref{eqn:snnlo} only depends on the combination
$\Upsilon_4 + \Upsilon_5$.  Similarly, \eqref{eqn:snnlo} only depends
on the combinations $\Upsilon_6+\Upsilon_8/2$ and
$\Upsilon_7+\Upsilon_8/2$, up to a term proportional to $\re\det(M
U^\dagger_0)$, which for $U_0 \in \text{SU}(N_f)$ reduces to $\re\det
M$ and can therefore be neglected for the same reason as for the term
proportional to $\h_2$.\footnote{For fixed topology we have $U_0 \in
  {\rm U}(N_f)$, and the contribution proportional to $\Upsilon_8 \re
  \det(M U_0^\dagger)$ in \eqref{eqn:snnlo} cannot be neglected.
  However, for two flavors $\y_8$ does not depend on the geometry
  \cite{Lehner:2010mv} and is proportional to $L_8$, which in turn is
  a combination of two-flavor LECs and HECs that are all of order
  $1/(4\pi)^2$ \cite{Gasser:1983yg}.}  To compute the $\Upsilon_i$ we
use eq.~(2.26) of \cite{Lehner:2010mv} and find
\begin{subequations}
  \label{eq:ups}
  \begin{align}
    \Upsilon_1 =-2\Upsilon_2=-2\Upsilon_3 &=
    \frac{P_4^r+4P_5^r}{2F^4V}=
    \begin{cases}
      -0.033 & \text{ for geometry } (a_2) \,, \\
      \phantom{-}0.0069 & \text{ for geometry } (b_2) \,,
    \end{cases} \\
    \Upsilon_4 + \Upsilon_5 &= -\frac{P_4^r+l_4^r}{2F^4V} = -0.017 \,,\\
    \Upsilon_6+\Upsilon_8/2 &=
    -\frac{\tfrac34P_4^r+l_3^r+l_4^r}{4F^4V} = -0.0074 \,, \\
    \Upsilon_7+\Upsilon_8/2 &= \frac{l_7^r}{4F^4V} = 0.0025 \,,
  \end{align}
\end{subequations}
see \cite{Lehner:2010mv} for the notation and the values of
$P_{4,5}^r$.  To arrive at the formulas for the $\Upsilon_i$ in
\eqref{eq:ups} we used the relation between the three-flavor NLO LECs
$L_i$ and the two-flavor NLO LECs $l_i$
\cite[eq.~(11.6)]{Gasser:1984gg}.  To compute the numbers, we took $F$
from \eqref{eq:Ffinal} and $l_7^r=l_7=0.005$
\cite[eq.~(19.21)]{Gasser:1983yg}.  For $l_3^r$ and $l_4^r$ we used
the running given in \cite[eq.~(10.18)]{Gasser:1983yg} with $M=m_\pi$
and $\mu=V^{-1/4}$ and took $\bar l_3=3.13$ and $\bar l_4=4.43$
\cite[table~X]{Allton:2008pn}. For $m_\pi$ we used the GOR relation,
resulting in $m_\pi=\sqrt{2m\Sigma}/F=110\text{ MeV}$, where we took
$\Sigma$ from \eqref{eq:Sigma}.  For the combinations of the $\y_i$ in
\eqref{eq:ups} the dependence on the renormalization scale drops out
so that the numbers given there are scale independent.

Note that in geometry $(b_2)$ all coefficients $\Upsilon_i$ are quite
small, which is consistent with our hypothesis that in this geometry
the systematic errors due to the non-universal terms are under
control.  In geometry $(a_2)$ the $\Upsilon_{1,2,3}$ are larger by a
factor of $\sim5$.  These terms appear to be the reason for the
systematic deviations observed in figure~\ref{fig:pda}.  As in the
case of $\Sigma$, the systematic errors can be reduced either by going
to a larger volume, or by computing the spectral correlation functions
corresponding to the effective action of eq.~\eqref{eqn:snnlo}
including all non-universal terms.  Again, this is beyond the scope of
this work.

To estimate the systematic errors on our determination of $\Sigma$ and
$F$, we recall that the NNLO finite-volume corrections and the
systematic deviations from RMT are of the same order in $1/F^2\sqrt
V$, see the comment after \eqref{eq:Sigma}.  The systematic errors due
to the NNLO finite-volume corrections can be estimated by comparing
the NLO and NNLO values given in \cite[table~2]{Lehner:2010mv}, and we
obtain $\approx 2\%$ for $F$ and $\approx 5\%$ for $\Sigma$.  Assuming
that for geometry $(b_2)$ the systematic errors due to non-universal
contributions in eq.~\eqref{eqn:snnlo} are roughly of the same
numerical size, we assign a total systematic error of $\approx 4\%$
to $F$ and of $\approx 10\%$ to $\Sigma$.

\section{Conclusions}\label{sec:conc}

We have shown that the geometry dependence of the Dirac eigenvalue
distributions at nonzero chemical potential strongly influences the
determination of $F$ from RMT fits.  Making a judicious choice of the
lattice geometry (in our case, by exchanging the temporal axis with
one of the spatial axes), this dependence can be significantly reduced
such that the systematic error on $F$ is kept under control.  This
makes the RMT-based method proposed in ref.~\cite{Damgaard:2005ys} a
useful alternative to other lattice methods.

Our final results for $\Sigma$ and $F$ obtained from the two-flavor
epsilon-regime run of JLQCD are given by
\begin{align}
  \Sigma^{\overline{\text{MS}}}(2 \text{ GeV}) &=
  Z_S^{\overline{\text{MS}}}(2 \text{ GeV})\ \Sigma = (230(1)(14)
  \text{ MeV})^3\,, & F &= 80(5)(5) \text{ MeV}\,,
\end{align}
where both values include finite-volume corrections at NLO and
$Z_S^{\overline{\text{MS}}}(2 \text{ GeV})=1.14(2)$
\cite{Fukaya:2007fb,Fukaya:2007yv}.  The left bracket gives the
statistical error, the right bracket gives the combined systematic
error (including the systematic error of the conversion to
$\overline{\text{MS}}$ and the uncertainty in the lattice spacing).
The individual systematic errors were added linearly.

To reduce the systematic errors due to the finite volume, it would be
beneficial to repeat this study at a larger simulation volume.  Also,
to eliminate the contamination caused by the non-universal terms, one
could attempt, within the framework of the finite-volume effective
theory of ref.~\cite{Lehner:2010mv}, to calculate Dirac eigenvalue
distributions beyond RMT including the systematic deviations at NNLO
in the epsilon expansion.  Work in this direction is in progress.

\acknowledgments 

This work was supported in part by BayEFG and the RIKEN FPR program
(CL), the Grant-in-Aid (No.~21674002) of the Japanese Ministry of
Education (SH), DFG grant SFB-TR 55 (JB and TW), and a KEK fellowship
(TW).  The numerical calculations were carried out on the IBM System
Blue Gene Solution at the High Energy Accelerator Research
Organization under support of its Large Scale Simulation Program
(No. 09/10-09).

\bibliographystyle{JHEP}
\bibliography{nnlonum}

\end{document}